\documentclass[twocolumn,twoside,slac_two]{revtex4}
\usepackage{graphicx}
\usepackage{fancyhdr}
\setcitestyle{sort&compress,numbers}

\begin{document}

\title{{\boldmath $C\!P$} violation in {\boldmath $D$} decays}

%

\author{V. Vagnoni} 
\affiliation{Istituto Nazionale di Fisica Nucleare, Sezione di Bologna, Italy}
\begin{abstract}
First evidence for $C\!P$ violation in two-body singly Cabibbo-suppressed decays of $D^0$ mesons reported by LHCb has recently aroused great interest in charm physics.
In this document the latest measurements of $C\!P$ violation in the charm sector are discussed. LHCb and CDF results on time-integrated $C\!P$ asymmetries in $D^0 \to \pi^-\pi^+$ and $D^0 \to K^-K^+$ decays are presented in some detail. A search for $C\!P$ violation performed by Belle in other two-body decays, namely $D^0 \to K^0_S \pi^0$, $D^0 \to K^0_S \eta^{(\prime)}$, $D^+_{(s)} \to \phi \pi^+$ and $D^+ \to \pi^+ \eta^{(\prime)}$, is also presented. Finally, results obtained by CDF with $D^0 \to K^0_S \pi^+\pi^-$ decays, as well as by LHCb and BaBar with other multi-body $D$ decays, are shown.
\end{abstract}

\maketitle

\thispagestyle{fancy}


\section{Introduction}
The violation of $C\!P$ symmetry is well established in the $K^0$ and $B^0$ meson systems~\cite{Christenson:1964fg,Aubert:2001nu,Abe:2001xe,Nakamura:2010zzi}, and first evidence in the $B^0_s$ system has been recently reported~\cite{LHCbBsACP}. Experimental measurements of $C\!P$ violation (CPV) in the quark flavour sector performed so far are generally well described by the Cabibbo-Kobayashi-Maskawa mechanism~\cite{Cabibbo:1963yz,Kobayashi:1973fv} of the Standard Model (SM). However, it is believed that the size of CPV in the SM is not sufficient to account for the asymmetry between matter and antimatter in the Universe \cite{Hou:2008xd}, hence additional sources of $C\!P$ symmetry breaking are being searched for as manifestations of physics beyond the SM.

The charm sector is a promising place to probe for the effects of physics beyond the SM. There has been a renaissance of interest in the past few years since evidence for $D^0$ mixing was first
seen~\cite{bib:babar_mixing_moriond,bib:belle_mixing_moriond}. Mixing is now well established~\cite{bib:hfag} at a level which is consistent with expectations~\cite{falk_grossman_ligeti_nir_petrov}. The recent evidence for CPV in singly Cabibbo-suppressed decays of $D^0$ mesons to two-body final states, reported by LHCb~\cite{CHARMCPVLHCB}, has definitely heightened the theoretical interest in charm physics. Prior to this measurement, $C\!P$ asymmetries in these decays were expected to be very small in the SM~\cite{bib:cicerone,bib:lenz,bib:grossman_kagan_nir,bib:petrov}, with
na\"\i ve predictions of up to $\mathcal{O}(10^{-3})$. For this reason, the asymmetry measured by LHCb, characterized by a central value of $\mathcal{O}(10^{-2})$, came as big surprise. 

Unfortunately, precise theoretical predictions of CPV in this sector are very difficult to achieve, as the charm quark is too heavy for chiral perturbation and too light for heavy-quark effective theory to be applied reliably. This fact means we cannot conclude that the observed effect is a clear sign of physics beyond the SM~\cite{theory1,theory2,theory3,theory4}. In order to investigate these promising hints further and clarify the picture, carrying out complementary measurements in other charm decays is of paramount importance.

In the next sections, after a description of the LHCb and CDF measurements with two-body $D^0$ decays, we discuss other recent measurements by the BaBar, Belle, CDF and LHCb collaborations, involving two-body as well as three- and four-body $D$ decays.

\section{{\boldmath $D^0 \to \pi^-\pi^+$} and {\boldmath $D^0 \to K^-K^+$}}

The time-dependent $C\!P$ asymmetry $A_{C\!P}(f;\,t)$ for $D^0$ decays to a self-conjugate $C\!P$ eigenstate $f$ 
(with $f = \bar{f}$) is defined as
\[
A_{C\!P}(f;\,t) = \frac{\Gamma(D^0(t) \to f)-\Gamma(\bar{D}^0(t) \to f)}{\Gamma(D^0(t) \to f)+\Gamma(\bar{D}^0(t) \to f)}, \label{eq:acpf}
\]
where $\Gamma$ is the decay rate for the process indicated.
In general $A_{C\!P}(f;\,t)$ depends on $f$.
For $f= K^- K^+$ and $f= \pi^- \pi^+$, $A_{C\!P}(f;\,t)$ can be expressed in terms of two contributions: 
  a direct component associated with CPV in the decay amplitudes, and
  an indirect component associated with CPV in the mixing or in the interference between mixing and decay.

The asymmetry $A_{C\!P}(f;\,t)$ may be written to first order as~\cite{bib:bigi_d2hh}
\[
A_{C\!P}(f;\,t) = a^{\mathrm{dir}}_{C\!P} (f) \, + \, \frac { t }{\tau} a^{\mathrm{ind}}_{C\!P}, \label{eq:acpphysicsth}
\]
where
$a^{\mathrm{dir}}_{C\!P} (f)$ is the direct $C\!P$ asymmetry,
$\tau$ is the $D^0$ lifetime, and 
$a^{\mathrm{ind}}_{C\!P}$ is the indirect $C\!P$ asymmetry, which is universal to a good approximation in the SM~\cite{Sokoloff}.
The time-integrated asymmetry measured by an experiment, $A_{C\!P}(f)$,
depends upon the time-acceptance of that experiment. It can be written as
\[
A_{C\!P}(f) = a^{\mathrm{dir}}_{C\!P} (f) \, + \, \frac {\langle t \rangle}{\tau} a^{\mathrm{ind}}_{C\!P}, \label{eq:acpphysics}
\]
where $\langle t \rangle$ is the average decay time in the reconstructed sample. Denoting by $\Delta$ the differences between quantities for $D^0 \to K^-K^+$ and $D^0 \to \pi^-\pi^+$ it is then possible to write
\begin{eqnarray}
\Delta A_{C\!P} & = & A_{C\!P}(K^-K^+) \, - \, A_{C\!P}(\pi^-\pi^+) \nonumber \\
&  = & \left[ a^{\mathrm{dir}}_{C\!P} (K^-K^+) \,-\, a^{\mathrm{dir}}_{C\!P} (\pi^-\pi^+) \right] \, + \, \frac {\Delta \langle t \rangle}{\tau} a^{\mathrm{ind}}_{C\!P}.\nonumber 
\end{eqnarray}
In the limit of vanishing  $\Delta \langle t \rangle$,
$\Delta A_{C\!P}$ becomes equal to the difference in the direct $C\!P$ asymmetry
between the two decays.
However, if the time-acceptance is different for the $K^-K^+$ and $\pi^-\pi^+$ final
states, a contribution from indirect CPV remains.

The LHCb collaboration performed a measurement of the difference in time-integrated $C\!P$ asymmetries between $D^0 \to K^-K^+$
and $D^0 \to \pi^-\pi^+$ using 0.62~fb$^{-1}$ of data~\cite{CHARMCPVLHCB}. The flavour of the initial state ($D^0$ or $\bar{D}^0$) is tagged by looking for the charge of the slow pion ($\pi_s^+$) in the decay chain $D^{*+} \to D^0 \pi_s^+$.

The raw asymmetry for tagged $D^0$ decays to a final state $f$
is given by $A_{\mathrm{raw}}(f)$, defined as
\begin{displaymath}
\resizebox{1\hsize}{!}{
$A_{\mathrm{raw}} (f) = \frac{N(D^{*+} \to D^0( f)\pi_s^+) \, - \, N(D^{*-} \to \bar{D}^0 (f)\pi_s^-)}
                                            {N(D^{*+} \to D^0( f)\pi_s^+) \, + \, N(D^{*-} \to \bar{D}^0 (f)\pi_s^-)},$
}
\end{displaymath}
where $N(X)$ refers to the number of reconstructed events of decay $X$
after background subtraction. 

To first order the raw asymmetries may be written as a sum of four components,
due to physics and detector effects:
\[
A_{\mathrm{raw}} (f) = A_{C\!P}(f) \, + \, A_\mathrm{D}(f) \, + \, A_\mathrm{D}(\pi_s^+) \, + \, A_\mathrm{P}(D^{*+}).
\]
Here, $A_\mathrm{D}(f)$ is the asymmetry in efficiency for the $D^0$ decay
into the final state $f$,  
$A_\mathrm{D}(\pi_s^+)$ is the asymmetry in efficiency for the slow pion
from the $D^{*+}$ decay chain, and 
$A_\mathrm{P}(D^{*+})$ is the production asymmetry
for $D^{*+}$ mesons.
The first-order expansion is valid since the individual asymmetries are small.

For a two-body decay of a spin-0 particle to a self-conjugate
final state there can be no $D^0$ detection asymmetry,
i.e. $A_\mathrm{D}(K^-K^+) = A_\mathrm{D}(\pi^-\pi^+) = 0.$  
Moreover, $A_\mathrm{D}(\pi_s^+)$ and $A_\mathrm{P}(D^{*+})$ are independent of $f$ and thus those terms cancel in the difference
$A_{\mathrm{raw}}  (K^-K^+) \, - \, A_{\mathrm{raw}}  (\pi^-\pi^+)$, resulting in
\[
\Delta A_{C\!P} = A_{\mathrm{raw}}  (K^-K^+) \, - \, A_{\mathrm{raw}}  (\pi^-\pi^+).\label{eq:adefequals}
\]

Note that the production asymmetry $A_\mathrm{P}(D^{*+})$ can be neglected in the case of the CDF experiment. This is because, in contrast with LHCb which is a forward spectrometer and employs flavour-asymmetric $pp$ collisions, CDF was a symmetric detector in pseudorapidity ($\eta$) and operated at a flavour-symmetric $p\bar{p}$ collider. Hence, integrating the measurement over a symmetric $\eta$ range, the possible presence of a production asymmetry is removed by construction, i.e. $A_\mathrm{P}(D^{*+})=0$.

\begin{figure}
\includegraphics[width=0.49\textwidth]{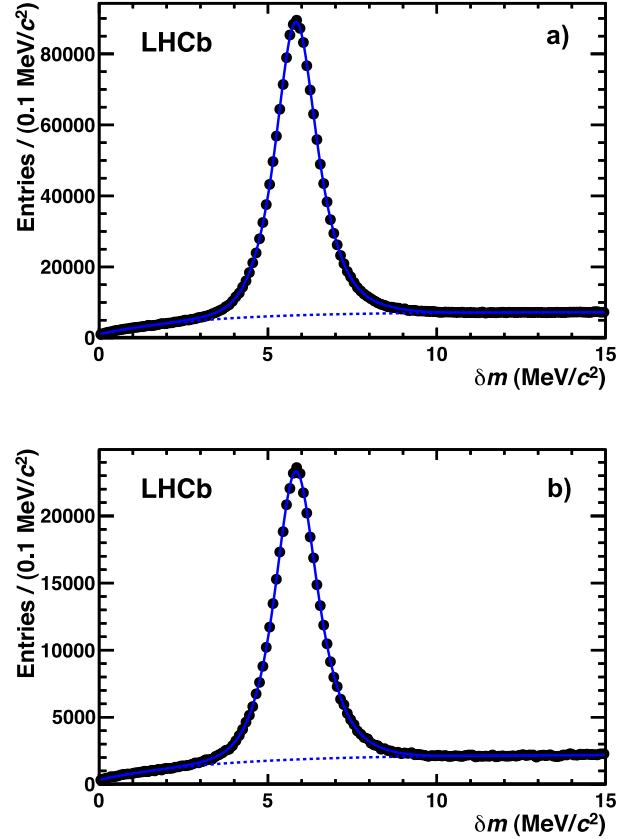}
\caption{LHCb fits to the $\delta m$ spectra, where the $D^0$ is reconstructed in the final states (a) $K^- K^+$ and (b) $\pi^- \pi^+$.}
\label{fig:deltaMassTagged}
\end{figure}

The mass difference ($\delta m$) spectra of candidates selected by LHCb as in Ref.~\cite{CHARMCPVLHCB}, where $\delta m = m(h^- h^+ \pi_s^+) - m(h^- h^+) - m(\pi^+)$
for $h=K,\pi$, are shown in Figure~\ref{fig:deltaMassTagged}.
The $D^{*+}$ signal yields are approximately
$1.44 \times 10^6$ in the $K^-K^+$ sample,
and $0.38 \times 10^6$ in the $\pi^-\pi^+$ sample.

Fits are performed on the samples in order to determine
$A_\mathrm{raw}(K^-K^+)$ and $A_\mathrm{raw}(\pi^-\pi^+)$.
The production and detection 
asymmetries can vary with $p_\mathrm{T}$ and pseudorapidity $\eta$, and so can the 
detection efficiency of the two different $D^0$ decays, in particular 
through the effects of the particle identification requirements. 
For this reason, since the different masses of the $K^-K^+$ and $\pi^-\pi^+$ final states lead to differences in the kinematic distributions of accepted signal events in the two cases, LHCb performs the analysis in several kinematic bins defined
by the $p_\mathrm{T}$ and $\eta$ of the $D^{*+}$ candidates,
the momentum of the slow pion, and the sign of $p_x$ of the
slow pion at the $D^{*+}$ vertex.

In each bin, one-dimensional unbinned maximum likelihood fits to the $\delta m$ spectra are performed.
A value of $\Delta A_{C\!P}$ is determined in each measurement bin
as the difference between $A_\mathrm{raw}(K^-K^+)$ and $A_\mathrm{raw} (\pi^-\pi^+)$.
A weighted average is performed to yield the result $\Delta A_{C\!P} =  (-0.82 \pm 0.21 )\%$,
where the uncertainty is statistical only.
Including the systematic uncertainty, LHCb measures the time-integrated difference in $C\!P$ asymmetry between $D^0 \to K^-K^+$ and $D^0 \to \pi^-\pi^+$ decays to be
\[
  \Delta A_{C\!P}^\mathrm{LHCb} = \left[ -0.82 \pm 0.21 (\mathrm{stat.}) \pm 0.11 (\mathrm{syst.}) \right]\%.
\]
Dividing the central value by the sum in quadrature of the
statistical and systematic uncertainties,
the significance of the measured deviation from zero is $3.5\sigma$. This is the first evidence for $C\!P$ violation in the charm sector.

\begin{figure}
\includegraphics[width=0.49\textwidth]{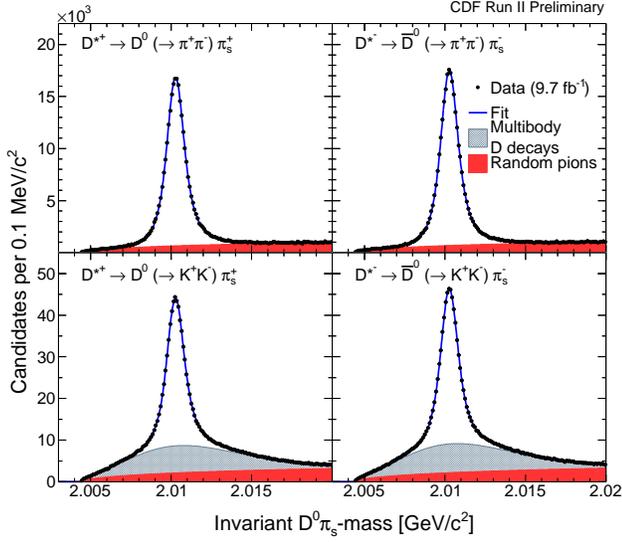}
\caption{CDF distributions of $D^0 \pi^+$ mass with fit results overlaid for (a) $D^0 \to \pi^-\pi^+$, (b) $\bar{D}^0 \to \pi^-\pi^+$, (c) $D^0 \to K^-K^+$ and (d) $\bar{D}^0 \to K^-K^+$.}
\label{fig:deltaMassTaggedCDF}
\end{figure}

A similar measurement has been performed by CDF~\cite{CDFCHARMCPV}. Figure~\ref{fig:deltaMassTaggedCDF} shows the invariant $D^0 \pi_s$ mass for $D^0$ decays to $\pi^-\pi^+$ and $K^-K^+$, corresponding to an integrated luminosity of 9.7~fb$^{-1}$. The $D^{*+}$ signal yields are approximately
$1.2 \times 10^6$ in the $K^-K^+$ sample,
and $0.55 \times 10^6$ in the $\pi^-\pi^+$ sample. These yields are very close to those of LHCb, despite a difference in integrated luminosity by a factor 15.

CDF uses a different approach from LHCb to take into account differences in kinematics between the two decay modes. The kinematic distributions are equalized by means of an event-by-event reweighting technique, and then a single fit is performed on the reweighted sample integrated in phase space.
The final result obtained by CDF is
\[
\Delta A_{C\!P}^\mathrm{CDF} = \left[ -0.62 \pm 0.21 (\mathrm{stat.}) \pm 0.10 (\mathrm{syst.}) \right]\%,
\]
which deviates from zero by $2.7\sigma$. This result is compatible with the LHCb measurement, with comparable accuracy and less than $1\sigma$ difference between the central values.

CDF has also provided individual measurements of $A_{C\!P}(D^0 \to K^-K^+)$ and $A_{C\!P}(D^0 \to \pi^-\pi^+)$~\cite{CDFCHARMCPVIND}, using a carefully constructed combination of raw asymmetries measured from tagged $D^{*+} \to D^0(K^-\pi^+)\pi^+$, $D^{*+} \to D^0(\pi^-\pi^+)\pi^+$, $D^{*+} \to D^0(K^-K^+)\pi^+$ and untagged $D^0 \to K^-\pi^+$ decays.
The calculation assumes that the production asymmetry of $D^{*+}$ is negligible. This holds at CDF but not at LHCb, where this kind of measurement is considerably more involved and has not been performed yet. Using part of the full data sample, corresponding to about 6~fb$^{-1}$, CDF obtains:
\begin{eqnarray}
A_{C\!P}^{D^0 \to K^-K^+} &=& \left[ -0.24 \pm 0.22 (\mathrm{stat.}) \pm 0.09 (\mathrm{syst.}) \right]\%,\nonumber\\
A_{C\!P}^{D^0 \to \pi^-\pi^+} &=& \left[ 0.22 \pm 0.24 (\mathrm{stat.}) \pm 0.11 (\mathrm{syst.}) \right]\%.\nonumber
\end{eqnarray}

\section{Other two-body decays}

\begin{figure}
\includegraphics[width=0.49\textwidth]{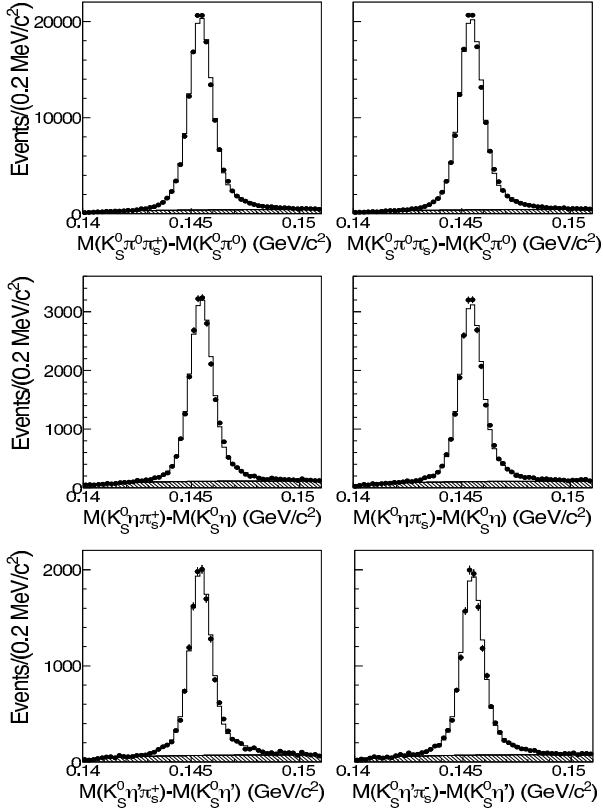}
\caption{Distributions of the mass difference $M(D^*)-M(D)$ for the $D^0 \to K^0_S P^0$ decay modes studied by Belle. Left plots show the mass difference between $D^{*+}$ and $D^0$ and right plots show that between $D^{*-}$ and $D^0$. The top plots are for the $K_S^0 \pi^0$ final state, the middle plots for $K_S^0\eta$, and the bottom plots for $K_S^0\eta^\prime$. The points with error bars are the data and the histograms show the results of the parameterizations of the data.}
\label{fig:belleksp0}
\end{figure}

The Belle collaboration has searched for CPV in the decay $D^0 \to K^0_S P^0$~\cite{BELLEKSP0}, where $P^0$ denotes a neutral pseudoscalar meson that is either a $\pi^0$, $\eta$, or $\eta^\prime$, using an integrated luminosity of 791~fb$^{-1}$. The observed $K^0_S P^0$ final states are mixtures of $D^0 \to \bar{K}^0 P^0$ and $D^0 \to K^0 P^0$ decays where the former are Cabibbo-favored and the latter are doubly Cabibbo-suppressed. SM $K^0 - \bar{K}^0$ mixing leads to a small $C\!P$ asymmetry in final states containing a neutral kaon, even if no $C\!P$ violating phase exists in the charm decay. Figure~\ref{fig:belleksp0} shows the distributions of the mass difference $M(D^*)-M(D)$ for the various decay modes. No evidence for CPV in these decays is observed, as Belle measures:
\begin{eqnarray}
A_{C\!P}^{D^0 \to K^0_S \pi^0}&=&\left[ -0.28 \pm 0.19 (\mathrm{stat.}) \pm 0.10 (\mathrm{syst.}) \right]\%,\nonumber\\
A_{C\!P}^{D^0 \to K^0_S \eta}&=&\left[ +0.54 \pm 0.51 (\mathrm{stat.}) \pm 0.16 (\mathrm{syst.}) \right]\%,\nonumber\\
A_{C\!P}^{D^0 \to K^0_S \eta^\prime}&=&\left[ +0.98 \pm 0.67 (\mathrm{stat.}) \pm 0.14 (\mathrm{syst.}) \right]\%.\nonumber
\end{eqnarray}

\begin{figure}
\includegraphics[width=0.49\textwidth]{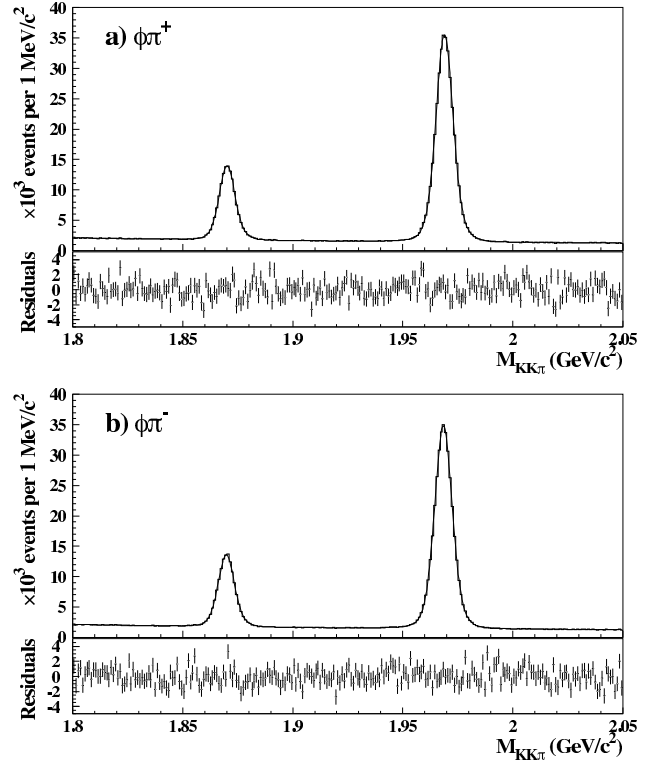}
\caption{Invariant mass distributions with the fitted functions superimposed for (a) $\phi \pi^+$ and (b) $\phi \pi^-$, observed by Belle.}
\label{fig:bellephipi}
\end{figure}

Belle has also searched for CPV in $D^+_{(s)} \to \phi \pi^+$~\cite{BELLEPHIPI} decays. For the $\phi \pi^+$, this is achieved by measuring the $C\!P$ violating asymmetries for the Cabibbo-suppressed decays $D^+ \to K^+K^-\pi^+$ and the Cabibbo-favored decays $D_s^+ \to K^+K^-\pi^+$ in the $K^+K^-$ mass region of the $\phi$ resonance, using 955~fb$^{-1}$ of data. The mass distributions are shown in Figure~\ref{fig:bellephipi}. Belle finds about $0.237 \times 10^6$ $D^\pm$ and $0.723 \times 10^6$ $D^\pm_s$ decays. Assuming negligible CPV in Cabibbo-favoured decays, Belle measures
\[
A_{C\!P}^{D^+ \to \phi \pi^+}=\left[ +0.51 \pm 0.28 (\mathrm{stat.}) \pm 0.05 (\mathrm{syst.})\right]\%.
\]
The result shows no evidence for CPV and agrees with SM predictions.

\begin{figure}
\includegraphics[width=0.49\textwidth]{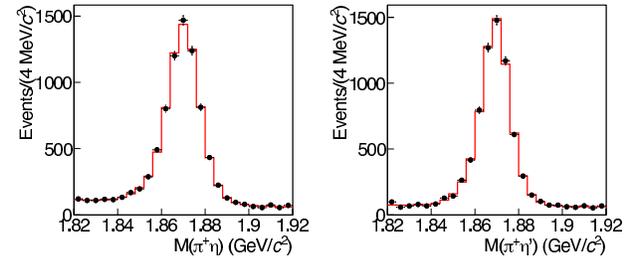}
\caption{Invariant mass distributions for (left) $\pi^+ \eta$ and (right) $\pi^+ \eta^\prime$ final states, observed by Belle. Points with error bars and histograms correspond to the data and the fit, respectively.}
\label{fig:bellepieta}
\end{figure}

Another relevant result has been obtained by Belle using 791~fb$^{-1}$ of data, with the most sensitive search for CPV in the decays $D^+ \to \pi^+ \eta$ and $D^+ \to \pi^+ \eta^\prime$~\cite{BELLEPIETA}. Figure~\ref{fig:bellepieta} shows the $\pi^+ \eta$ and $\pi^+ \eta^\prime$ invariant mass distributions. The final results obtained by Belle for CPV in these decay modes are:
\begin{eqnarray}
A_{C\!P}^{D^+ \to \pi^+ \eta}&=&\left[ +1.74 \pm 1.13 (\mathrm{stat.}) \pm 0.19 (\mathrm{syst.})\right]\%,\nonumber\\
A_{C\!P}^{D^+ \to \pi^+ \eta^\prime}&=&\left[ -0.12 \pm 1.12 (\mathrm{stat.}) \pm 0.17 (\mathrm{syst.})\right]\%.\nonumber
\end{eqnarray}
Again, no evidence for CPV is found with these modes.

\section{{\boldmath $D^0 \to K^0_S \pi^+\pi^-$}}

\begin{figure}
\includegraphics[width=0.49\textwidth]{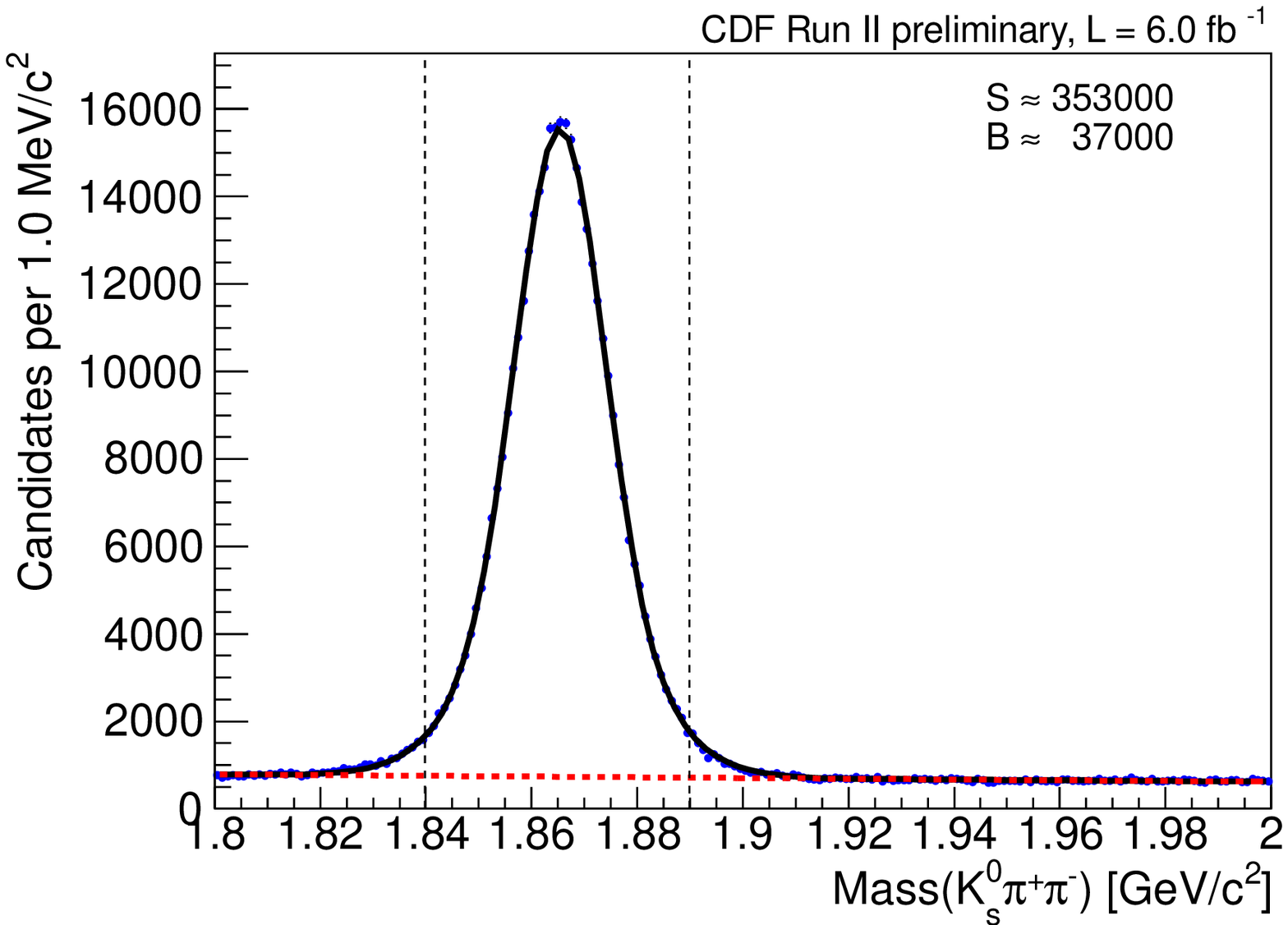}
\includegraphics[width=0.49\textwidth]{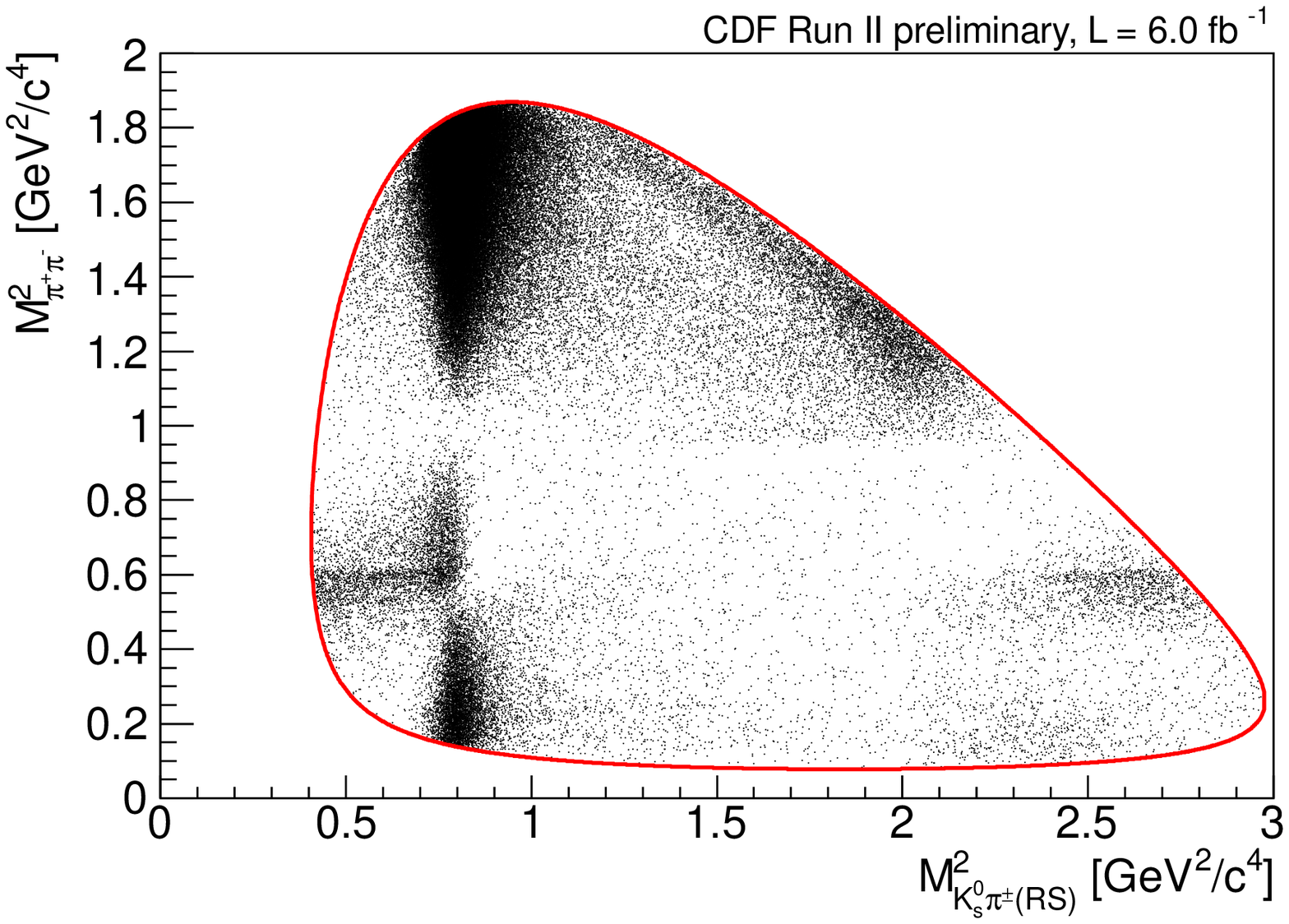}
\caption{Invariant $K^0_S \pi^+\pi^-$ mass distribution (top) and $D^0 \to K^0_S \pi^+\pi^-$ Daltiz plot (bottom) obtained by CDF.}
\label{fig:CDFKSPIPI}
\end{figure}

With the same trigger used to collect $D^0 \to h^+h^-$ decays, and using an offline selection based on a neural network, CDF is able to reconstruct about $0.35 \times 10^6$ decays to resonances of $D^*$ tagged decays, using 6~fb$^{-1}$ of integrated luminosity~\cite{CDFKSPIPI}. Figure~\ref{fig:CDFKSPIPI} shows the $K^0_S \pi^+\pi^-$ mass spectrum and the $D^0 \to K^0_S \pi^+\pi^-$ Dalitz plot obtained by CDF. The analysis is made both in a model-independent way, by binning the Dalitz plot and looking for bin-by-bin asymmetries, and by fitting the population of each resonance in $D^0$ and $\bar{D}^0$ decays using an isobar model. 

The search for CPV in bin-by-bin asymmetries shows no evidence of deviation from a Gaussian distribution with zero mean and unit width. The full Dalitz fit includes a parameterization of the efficiency over the Dalitz plane. An event-by-event reweighting is applied in order to equalize $D^{*+}$ and $D^{*-}$ kinematics. Again, no evidence of CPV is found in any of the considered resonant modes. CDF measures the following asymmetry integrated over the Dalitz plane:
\[
  A_{C\!P} = \left[ -0.05 \pm 0.57 (\mathrm{stat.}) \pm 0.54 (\mathrm{syst.}) \right]\%.
\]

\section{Other multi-body {\boldmath $D$} decays}

\begin{figure}
\includegraphics[width=0.49\textwidth]{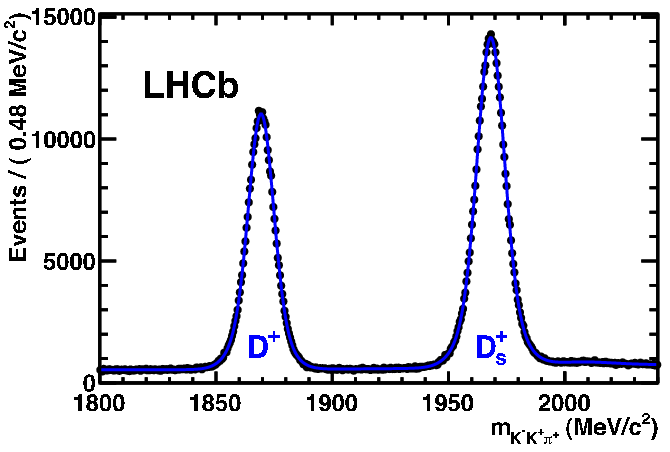}
\includegraphics[width=0.49\textwidth]{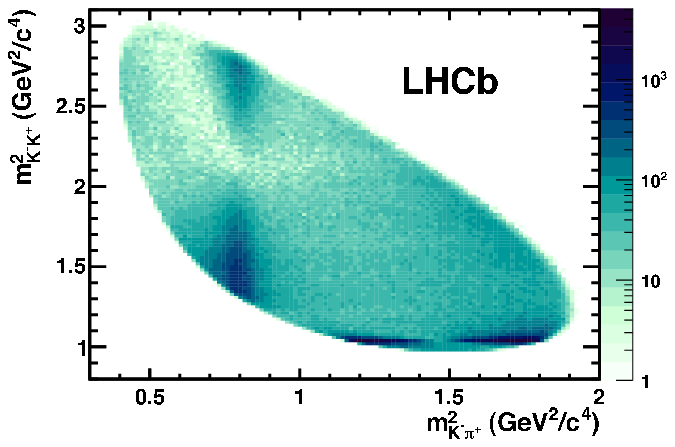}
\caption{Invariant $K^-K^+\pi^+$ mass distribution (top) and $D^+ \to K^-K^+\pi^+$ Daltiz plot (bottom) obtained by LHCb.}
\label{fig:lhcb3body}
\end{figure}

LHCb has demonstrated the ability to select large samples of three-body decays. For example, about $0.37 \times 10^6$ signal $D^+ \to K^-K^+\pi^+$ decays with high purity have been selected using 35~pb$^{-1}$ of integrated luminosity with data taken during 2010~\cite{LHCBTHREEBODY}. The reconstructed $K^-K^+\pi^+$ mass distribution and the $D^+ \to K^-K^+\pi^+$ Dalitz plot are shown in Figure~\ref{fig:lhcb3body}. The charge asymmetries in the control modes $D^+ \to K^-\pi^+\pi^+$ and $D^+_s \to K^-K^+\pi^+$ are investigated to eliminate the possibility
of observing detector asymmetries as signals of CPV. No significant charge asymmetries are observed, indicating that such systematic effects are negligible at this level of precision. The Dalitz plot of the signal mode is then binned according to four binning schemes, two of which
account for the resonant structure of the decay and two of which do not. The distributions of bin-by-bin asymmetries are consistent with no CPV in each of the binning schemes.

With four-body decays the phase space becomes five-dimensional, and the difficulty of the analysis grows considerably. However, in four-body decays the method of $T$-odd moments becomes available. This has been applied e.g. by the BaBar collaboration searching for CPV in $D^+ \to K^0_S K^+ \pi^+ \pi^-$ decays~\cite{BABARTODD}. The procedure consists in defining a triple product of the momenta of three of the particles, i.e. $C_T = \vec{p}_{K^+} \cdot (\vec{p}_{\pi^+} \times \vec{p}_{\pi^-})$. It is then possible to define the quantity $A_T$ for $D^+$ decays (and its analogue $\bar{A}_T$ for $D^-$ decays) as:
\[
A_T = \frac{\Gamma(C_T >0) - \Gamma(C_T<0)}{\Gamma(C_T >0) + \Gamma(C_T<0)}.
\]
Then, if the quantity $\mathcal{A}_T=\frac{1}{2}(A_T-\bar{A}_T)$ differs from zero, this is a sign of CPV. No evidence for CPV is found, as BaBar obtains:
\[
\mathcal{A}_T=\left[ -1.20 \pm 1.00 (\mathrm{stat.}) \pm 0.46 (\mathrm{syst.}) \right]\%.
\]

\section{Conclusions}

We have summarized some of the latest results involving two-, three- and four-body $D$ decays at BaBar, Belle, CDF and LHCb. In the past months, first evidence of CPV in charm decays has been first obtained by LHCb with a $3.5\sigma$ significance, then followed by CDF with $2.7\sigma$. This has triggered a big effort by the theory community, in order to understand whether any effect of physics beyond the SM was manifesting itself or whether the result could be explained in the framework of the SM. Precise theoretical predictions in this sector are very difficult to achieve unfortunately, so it is not yet possible to draw firm conclusions. Nevertheless, if the result is confirmed at more than $5\sigma$ with increased statistics by LHCb, one expects CPV to show up in other charm decays as well, and the pattern of asymmetries will help theorists to decode the overall picture and determine whether it is truly a SM effect. For the moment, all the efforts looking for CPV in other $D$ decays besides $D^0 \to \pi^+\pi^-$ and $D^0 \to K^+K^-$ have been frustrated. But the search has just started.

\bigskip 

\begin{thebibliography}{9}   

\bibitem{Christenson:1964fg}
J.~H. Christenson, J.~W. Cronin, V.~L. Fitch, and R.~Turlay,
Phys. Rev. Lett. {\bf 13} (1964) 138.

\bibitem{Aubert:2001nu}
BaBar collaboration, B.~Aubert {\em et~al.},
Phys. Rev. Lett. {\bf 87} (2001) 091801.

\bibitem{Abe:2001xe}
Belle collaboration, K.~Abe {\em et~al.},
Phys. Rev. Lett. {\bf 87} (2001) 091802.

\bibitem{Nakamura:2010zzi}
Particle Data Group, K.~Nakamura {\em et~al.},
J. Phys. {\bf G37} (2010) 075021.

\bibitem{LHCbBsACP}
LHCb collaboration, R.~Aaij {\em et~al.},
Phys. Rev. Lett.  {\bf 108} (2012) 201601.

\bibitem{Cabibbo:1963yz}
N.~Cabibbo,
Phys. Rev. Lett. {\bf 10} (1963) 531.

\bibitem{Kobayashi:1973fv}
M.~Kobayashi and T.~Maskawa,
Prog. Theor. Phys. {\bf 49} (1973) 652.

\bibitem{Hou:2008xd}
W.-S. Hou,
Chin. J. Phys. {\bf 47} (2009) 134.

\bibitem{bib:babar_mixing_moriond}
BaBar collaboration, B.~Aubert {\em et~al.},
Phys. Rev. Lett. {\bf 98} (2007) 211802.

\bibitem{bib:belle_mixing_moriond}
Belle collaboration, M.~Staric {\em et~al.},
Phys. Rev. Lett. {\bf 98} (2007) 211803.

\bibitem{bib:hfag}
Heavy Flavor Averaging Group, D.~Asner {\em et~al.},
arXiv:1010.1589 [hep-ex].

\bibitem{falk_grossman_ligeti_nir_petrov}
A.~F. Falk, Y.~Grossman, Z.~Ligeti, Y.~Nir, and A.~A. Petrov,
Phys. Rev. D {\bf 69} (2004) 114021.

\bibitem{CHARMCPVLHCB}
LHCb collaboration, R.~Aaij {\it et al.},
Phys.\ Rev.\ Lett.\  {\bf 108} (2012) 111602.

\bibitem{bib:cicerone}
S.~Bianco, F.~L. Fabbri, D.~Benson, and I.~Bigi,
Riv. Nuovo Cim. {\bf 26N7} (2003) 1.

\bibitem{bib:lenz}
M.~Bobrowski, A.~Lenz, J.~Riedl, and J.~Rohrwild,
JHEP {\bf 1003} (2010) 009.

\bibitem{bib:grossman_kagan_nir}
Y.~Grossman, A.~L. Kagan, and Y.~Nir,
Phys. Rev. D {\bf 75} (2007) 036008.

\bibitem{bib:petrov}
A.~A. Petrov,
PoS {\bf BEAUTY2009} (2009) 024.

\bibitem{theory1}
E.~Franco, S.~Mishima and L.~Silvestrini,
JHEP {\bf 1205} (2012) 140.

\bibitem{theory2}
  G.~F.~Giudice, G.~Isidori and P.~Paradisi,
JHEP {\bf 1204} (2012) 060.

\bibitem{theory3}
Y.~Grossman, A.~L.~Kagan and J.~Zupan,
Phys.\ Rev.\ D {\bf 85} (2012) 114036.

\bibitem{theory4}
B.~Bhattacharya, M.~Gronau and J.~L.~Rosner,
arXiv:1207.0761 [hep-ph].

\bibitem{bib:bigi_d2hh}
I.~I. Bigi, A.~Paul, and S.~Recksiegel,
JHEP {\bf 1106} (2011) 089.

\bibitem{Sokoloff}
A.~L. Kagan and M.~D. Sokoloff,
Phys. Rev. D {\bf 80} (2009) 076008.

\bibitem{CDFCHARMCPV}
CDF collaboration, T.~Aaltonen {\it et al.}, 
arXiv:1207.2158 [hep-ex].

\bibitem{CDFCHARMCPVIND}
CDF collaboration, T.~Aaltonen {\em et~al.},
Phys. Rev. D {\bf 85} (2012) 012009.

\bibitem{BELLEKSP0}
Belle collaboration, B.~R.~Ko {\it et al.},
Phys.\ Rev.\ Lett.\  {\bf 106} (2011) 211801.

\bibitem{BELLEPHIPI}
Belle collaboration, M.~Staric {\it et al.},
Phys.\ Rev.\ Lett.\  {\bf 108} (2012) 071801.

\bibitem{BELLEPIETA}
Belle collaboration, E.~Won {\it et al.},
Phys.\ Rev.\ Lett.\  {\bf 107} (2011) 221801.

\bibitem{CDFKSPIPI}
CDF collaboration, T.~Aaltonen {\it et al.},
Phys.\ Rev.\ D {\bf 86} (2012) 032007.

\bibitem{LHCBTHREEBODY}
LHCb collaboration, R.~Aaij {\it et al.},
Phys.\ Rev.\ D {\bf 84} (2011) 112008.

\bibitem{BABARTODD}
BaBar collaboration, J.~P.~Lees {\it et al.},
Phys.\ Rev.\ D {\bf 84} (2011) 031103.

\end{thebibliography}

\end{document}